\documentclass[a4paper,fleqn,usenatbib]{mnras}

\usepackage[T1]{fontenc}
\usepackage{ae,aecompl}

\usepackage{graphicx}	
\usepackage{amsmath}	
\usepackage{amssymb}	

\title[Compton thick absorber in 3C~345]{Compton thick absorber in type 1 quasar 3C~345 revealed by \textit{Suzaku} and \textit{Swift}/BAT}

\author[S. Eguchi]{
Satoshi Eguchi,$^{1}$\thanks{E-mail: satoshieguchi@fukuoka-u.ac.jp}
\\
$^{1}$Department of Applied Physics, Faculty of Science, Fukuoka University, 8--19--1, Nanakuma, Jonan-ku, Fukuoka 814--0180, Japan
}

\date{Accepted XXX. Received YYY; in original form ZZZ}

\pubyear{2017}

\begin{document}
\label{firstpage}
\pagerange{\pageref{firstpage}--\pageref{lastpage}}
\maketitle

\begin{abstract}
The archival data of 3C~345, a type 1 quasar at z = 0.5928,
obtained with \textit{Suzaku} and \textit{Swift}/BAT are analysed.
Though previous studies of this source applied only a simple broken
power law model, a heavily obscuring material is found to be required
by considering Akaike information criteria.
The application of the numerical torus model by \cite{Murphy2009}
surprisingly reveals the existence of Compton thick type 2
nucleus with the line-of-sight hydrogen column density of
the torus of $N_{\textrm{H}} = 10^{24.5} \ \textrm{cm}^{-2}$
and the inclination angle of $\theta_{\textrm{inc}} = 90\degr$.
However, this model fails to account for the Eddington ratio
obtained with the optical observations by \citet{Gu2001}
and \citet{Shen2011}, or requires the existence of
a supermassive black hole binary, which was suggested by
\cite{Lobanov2005}, thus this model is likely to be inappropriate
for 3C~345.
A partial covering ionized absorber model which accounts
for absorption in ``hard excess'' type 1 AGNs is
also applied, and finds a Compton thick absorber with
the column density of $N_{\textrm{H}} \simeq 10^{25} \ \textrm{cm}^{-2}$,
the ionization parameter of $\log \xi \gtrsim 2$, and
the covering fraction of $75\% \lesssim f_{c} \lesssim 85\%$.
Since this model obtains a black hole mass of $\log \left( M_{\textrm{BH}} / M_{\sun} \right) = 9.8$,
which is consistent with the optical observation by \citet{Gu2001},
this model is likely to be the best-fitting model of this source.
The results suggest that 3C~345 is the most distant and most
obscured hard excess AGN at this time.
\end{abstract}




\section{Introduction}
An active galactic nucleus (AGN) is one of the most energetic phenomena
in the universe.
AGNs are essentially classified into two types (type 1 and 2) based on
whether or not broad lines such as \ion{H}{I}, \ion{He}{I}, and \ion{He}{II}
are observed.
According to unified models of AGNs \citep[e.g.,][]{Urry1995}, this difference
is due to the viewing angle of a dust torus, which surrounds a central
supermassive black hole, accretion disc, and a region emitting broad lines
(broad line region; BLR), since light from the BLR is blocked by the
torus if viewed from an edge-on angle.
This picture is supported by the findings of hidden polarized broad lines
in the optical spectra of type 2 AGNs \citep[e.g.,][]{Antonucci1985}.

Studies based on population synthesis models of the cosmic X-ray background
(CXB) suggest that a significant fraction of type 2 AGNs
have a line-of-sight hydrogen column density of
$N_{\rm H} \gtrsim 10^{24} \ \rm{cm}^{-2}$ \citep[e.g.,][]{Gilli2007,Ueda2014},
where the torus is optically thick for Compton scattering
(Compton thick; CT).
Numerical simulations predict that most AGNs experience such heavily
obscured phase in their early stage of growth \citep[e.g.,][]{Hopkins2005}.
To study these objects provides us fundamental information of the cosmic
evolution of supermassive black holes and galaxies.
On the other hand, recent X-ray observations find that there
are CT absorbers in some type 1 AGNs \citep[e.g.,][]{Turner2009}.
Hence the hydrogen column density is one of key parameters to
characterize an AGN as well as the black hole mass and accretion rate.
However, these sources are missed by observations below 10 keV
due to the strong photoelectric absorption.
Sensitive hard X-ray observations above 10 keV, where the penetrating
power overwhelms the absorption, are crucial for these sources.

3C~345 ($z = 0.5928$) is a type 1 quasar in the 3C catalogue
\citep{Edge1959} referred to as core-dominated radio source since
its high-frequency radio emission is dominated by a compact flat
spectrum \citep[][and references therein]{Laing1983}.
It was firstly detected as a 2-keV X-ray source by \textit{Einstein} \citep{Ku1980}.
\citet{Neugebauer1979} found that the optical and infrared continua
of the source show strong time variability on time scales of months.
\citet{Moore1981} observed strong polarization and its large changes
occurred on time scales of a week in the optical band.
The apparent velocity of the jet component of the source is
superluminal ($v \simeq 15 c$) \citep{Unwin1983}.
The supermassive black hole mass is estimated to be
$\log \left( M_{\textrm{BH}} / M_{\sun} \right) = 9.901$
based on the H$\beta$ line width by \citet[][hereafter GCJ01]{Gu2001}.
Similarly, \citet[][hereafter S11]{Shen2011} derived the black hole mass
of $\log \left( M_{\textrm{BH}} / M_{\sun} \right) = 9.27 \pm 0.09$
and the Eddington ratio of $\log \lambda_{\textrm{Edd}} = -0.1$
based on the H$\beta$ line width and its luminosity.
Recently, observations with \textit{Chandra} and
\textit{Hubble Space Telescope} were performed \citep{Kharb2012}
to constrain the physical properties of the jet.

While 3C~345 is intensively studied for its unique natures,
there is no report of the existence of a CT absorber at this time;
I analysed archival data of 3C~345 obtained with \textit{Suzaku}
and \textit{Swift}/BAT by utilizing the numerical torus model
provided by \citet[][hereafter MY09]{Murphy2009}\footnote{\url{http://www.mytorus.com/}}
and photo-ionization models computed with the XSTAR code
\citep{Kallman2001}, and found that this source is obscured
by a CT material.
In this paper, I present the results of a detailed analysis of the \textit{Suzaku}
and \textit{Swift}/BAT spectra.
This paper is organized as follows.
Section \ref{section-observation} describes the observations
and data reduction.
Firstly, I analyse the spectra with a conventional broken power
law model in Section \ref{section-broken-powerlaw}.
Next, the spectra are analysed with torus absorption models
in Section \ref{section-torus-absorption}.
Lastly, I present the results obtained with partial covering
absorber models in Section \ref{section-partial-covering}.
The discussion and summary follow Section \ref{section-discussion} and \ref{section-summary}.
I adopt the cosmological parameters $(H_{0}, \Omega_{\textrm{m}}, \Omega_{\lambda})
= (70 \ \textrm{km} \ \textrm{s}^{-1} \ \textrm{Mpc}^{-1}, 0.3, 0.7)$,
the photoelectric absorption cross-sections of \citet{Verner1996}
(\texttt{vern} in XSPEC), and the solar abundances of \citet{Anders1989}
(\texttt{angr} in XSPEC) through the paper.
The errors are 90\% confidence limits for a single parameter.

\section{Observation and Data Reduction} \label{section-observation}

\subsection{Observation}

\textit{Suzaku} observed 3C~345 on 2012 September 11 with a net exposure
of 12.7 ks.
The data are public on the \textit{Suzaku} page on Data ARchives and Transmission
System (DARTS)\footnote{\url{http://www.darts.isas.jaxa.jp/astro/suzaku/data.html}},
and the observation ID is 707043010.
\textit{Suzaku} \citep{Mitsuda2007} carries four X-ray CCD cameras called the X-ray
Image Spectrometers (XIS-0, XIS-1, XIS-2, and XIS-3), which cover the 0.2--12 keV
band, as the focal plane imager of four X-ray telescopes, and non-imaging
instrument called the Hard X-ray Detector (HXD) consisting of Si PIN
photo-diodes and Gadolinium Silicon Oxide (GSO) scintillation counters,
which cover the 10--70 keV and 40--600 keV band, respectively.
XIS-0, XIS-2, and XIS-3 are front-side illuminated CCDs (FI-XISs),
and XIS-1 is the back-side illuminated one (BI-XIS).
Since XIS-2 became inoperable on 2007 November 7 \citep{Dotani2007},
no XIS-2 data is available.
Spaced-row charge injection (SCI) was applied to the XIS data to improve
the energy resolution \citep{Nakajima2008}.
3C~345 was observed at the XIS nominal position.
In the spectral analysis, the 70-month (between 2004 December and 2010 September)
integrated \textit{Swift}/BAT spectrum covering the 15--200 keV band
\citep[Swift J1643.1+3951,][]{Baumgartner2013}\footnote{\url{http://swift.gsfc.nasa.gov/results/bs70mon/}}
is utilized.

\subsection{Data Reduction} \label{subsection-data-reduction}

The \textit{Suzaku} data are reduced with the HEAsoft version 6.18 and the
latest version of CALDB on 2016 February 24.
All event files are reprocessed with the \texttt{aepipeline} command, and
the produced cleaned events are analysed.
The light curves and spectra of XISs are extracted from a circular region
with a $1\arcmin.5$-radius around the detected position.
The backgrounds are taken from a circular source-free region with a
$3\arcmin$-radius.
The so-called ``tuned'' non-X-ray background (NXB) model provided by
the HXD team is used for the HXD/PIN data, whose systematic errors
are estimated to be $\simeq 1.4\%$ at a $1\sigma$ confidence level
in the 15--40 keV band for a 10 ks exposure \citep{Fukazawa2009}.
The CXB spectrum simulated with the HXD/PIN response for a uniformly extended
emission is added to the NXB spectrum.
The HXD/GSO data are not analysable since no background model is provided.

\subsection{Light Curves}

\begin{figure*}
 \centerline{
  \includegraphics[width=0.9\columnwidth]{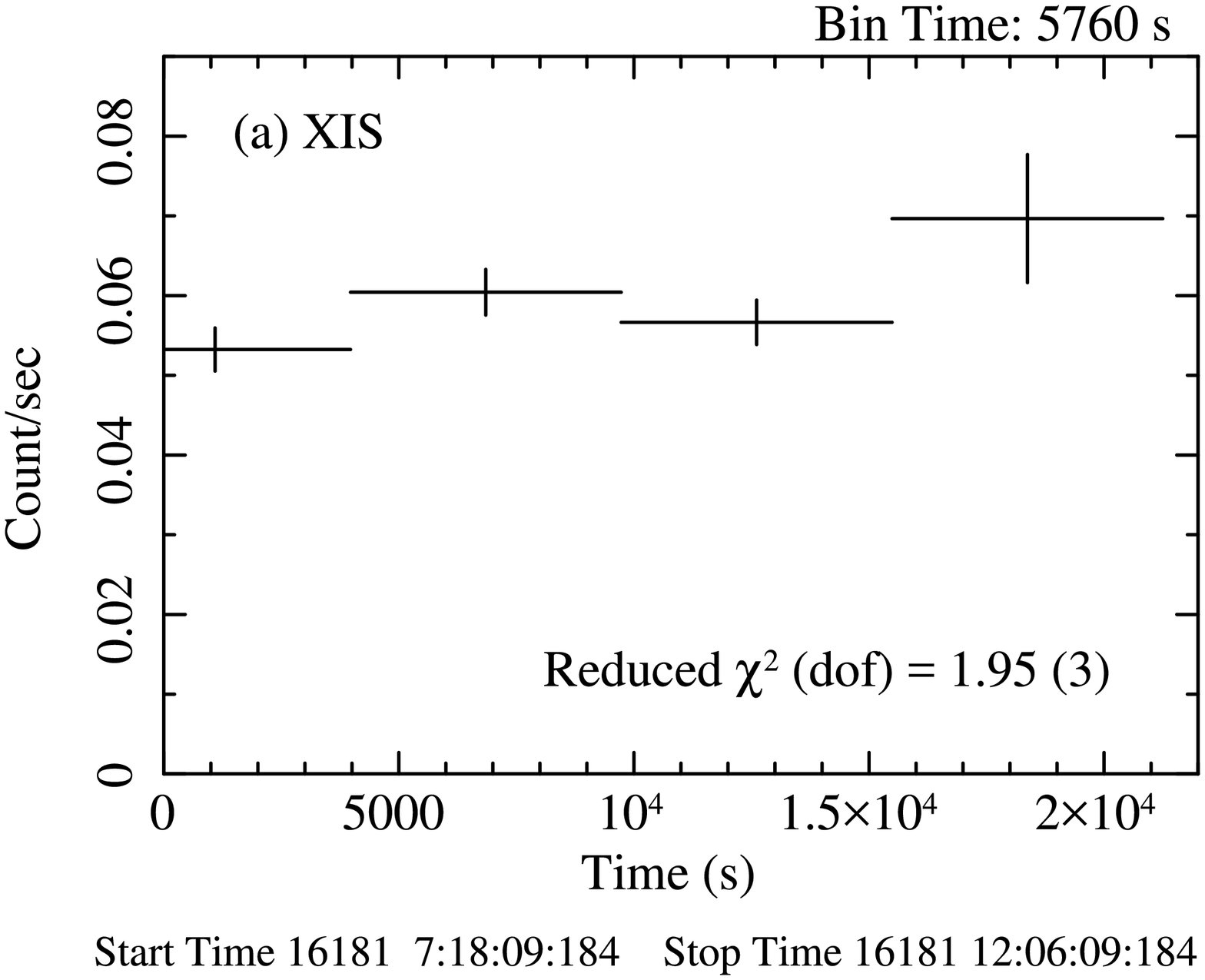}
  \hspace{0.1\columnwidth}
  \includegraphics[width=0.9\columnwidth]{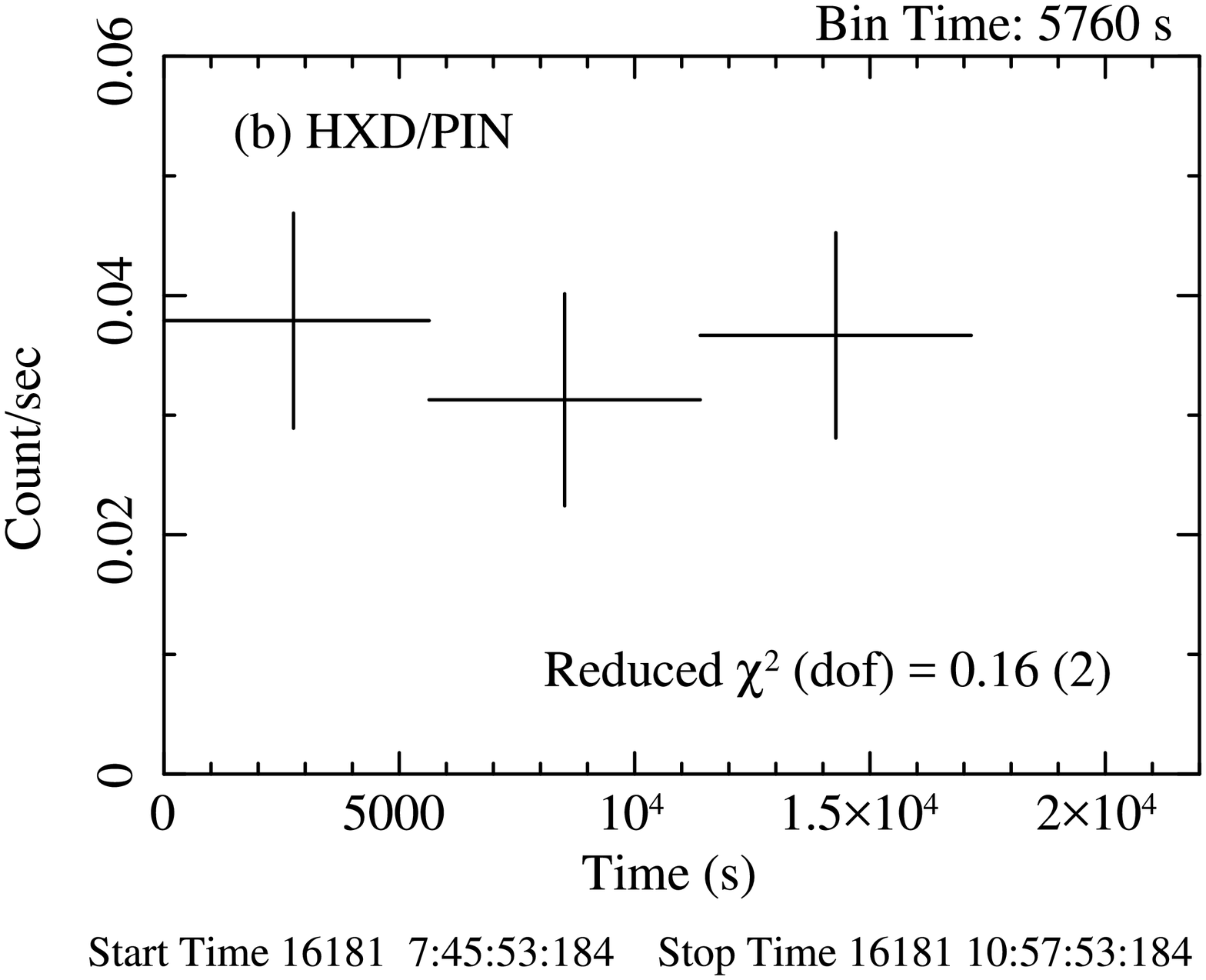}
 }
 \caption{
  Background-subtracted light curves of Suzaku.
  One bin corresponds to 96 minutes.
  The numbers listed in each panel are the value of reduced $\chi^{2}$ with
  the degrees of freedom for the constant flux hypothesis.
  Left: the light curves of the XIS in the 2--10 keV band,
  where the data from XIS-0 and XIS-3 are summed.
  Right: the light curves of the HXD/PIN in the 15--40 keV band.
  Note that the HXD/PIN observation started about 30 minutes after
  the beginning of the XIS observation.
 }
 \label{figure-light-curve}
\end{figure*}

Figure~\ref{figure-light-curve} shows the background-subtracted light curves
of 3C~345 obtained with the \textit{Suzaku} XIS and HXD/PIN in the 2--10 keV
and 15--40 keV band, respectively.
The data from XIS-0 and XIS-3 are summed.
To minimize any systematic uncertainties caused by the orbital change of
the satellite, the data taken during one orbit ($\simeq$ 96 minutes) are
merged into one bin, and this yields 4 and 3 bins for XIS and HXD/PIN,
respectively;
note that the HXD/PIN observation started about 30 minutes after
the XIS observation started according to the FITS headers.
To check whether there are any significant time variabilities during
the observation, I perform a simple $\chi^{2}$ test to each light curve,
assuming a null hypothesis of constant flux.
The resultant reduced $\chi^{2}$ value and the degrees of freedom are
superimposed on Figure~\ref{figure-light-curve}.
Though the time variability in the XIS cannot be rejected at the 90\%
confidence level, the flux changes are marginal;
the one in the HXD/PIN can be ruled out at the level.
Thus the time-averaged spectra over the entire observation are analysed.

\subsection{Spectra}

The spectra of FI-XISs are summed;
the data of the FI-XISs, BI-XIS, HXD/PIN, and \textit{Swift}/BAT
in the energy band of 0.8--10.0 keV, 0.5--8.0 keV, 15-25 keV, and 15--100 keV,
respectively, are used, covering the 0.5--100 keV band simultaneously.
The relative normalization of the PIN with respect to the FI-XISs
is fixed at 1.16 based on the calibration of the Crab Nebula \citep{Maeda2008}.
Those of BI-XIS and BAT with respect to the FI-XISs are set as free parameters.
Galactic absorption is always included in all the models discussed
in this paper;
its hydrogen column density is fixed at the value calculated with the \texttt{nh}
command, $N_{\textrm{H}}^{\textrm{Gal}} = 1.14 \times 10^{20} \ \textrm{cm}^{-2}$,
which is based on the result of the \ion{H}{I} map \citep{Kalberla2005}.

\section{Conventional Broken Power Law Model} \label{section-broken-powerlaw}

\begin{table*}
 \centering
 \caption{
  Best-fitting parameters with Model A 
 }
 \label{table-model-a}
 \begin{tabular}{ccccc}
  \hline
   & Energy Band & 0.5--100 keV & 0.5--6.0 keV & 6.0--100 keV \\
  \hline
  (1) & $N_{\rm H}$ ($10^{24} \ \textrm{cm}^{-2}$) & $2.3 \times 10^{-11}$ ($< 4.4 \times 10^{-3}$) & $2.0 \times 10^{-3}$ ($< 5.0 \times 10^{-3}$) & $1.2 \pm 0.7$ \\
  (2) & $\Gamma_{\textrm{soft}}$ & $1.67 \pm 0.07$ & $2.3^{+0.9}_{-0.4}$ &  1.7$^{a}$ \\
  (3) & $\Gamma_{\textrm{hard}}$ & $1.1^{+0.5}_{-0.4}$ & $1.7 \pm 0.1$ & $1.3 \pm 0.4$ \\
  (4) & $E_{\textrm{break}}$ (keV) & $8^{+10}_{-6}$ & $2.1^{+0.9}_{-0.4}$ & $10$ ($< 45$) \\
  (5) & $A_{\textrm{BAT}}$ & $0.6^{+1.0}_{-0.4}$ & --- & $0.40 \pm 0.07$ \\
  (6) & $F_{\textrm{2-10}}$ ($10^{-12} \ \textrm{erg} \ \textrm{cm}^{-2} \ \textrm{s}^{-1}$) & 2.7 & 2.6 & 1.4 \\
  (7) & $F_{\textrm{10-50}}$ ($10^{-11} \ \textrm{erg} \ \textrm{cm}^{-2} \ \textrm{s}^{-1}$) & 0.98 & 0.45 & 1.7 \\
  (8) & $L_{\textrm{2-10}}$ ($10^{46} \ \textrm{erg} \ \textrm{s}^{-1}$) & 0.32 & 0.33 & 0.83 \\
  (9) & $\textrm{AIC}_{c}$ & $-77.42$ & --- & --- \\
   & $\chi^{2} / \textrm{d.o.f.}$ & $111.57 / 177$ & $90.75 / 158$ & $6.16 / 12$ \\
  \hline
 \end{tabular}
 \begin{flushleft}
  $^{a}$Fixed.
  \\
  Note.
  (1) The line-of-sight hydrogen column density.
  (2) The power law photon index below $E_{\textrm{break}}$.
  (3) The power law photon index above $E_{\textrm{break}}$.
  (4) The break energy.
  (5) The relative normalization of the BAT with respect to the FI-XISs.
  (6) The observed flux in the 2--10 keV band.
  (7) The observed flux in the 10--50 keV band.
  (8) The 2--10 keV intrinsic luminosity corrected for the absorption.
  (7) The corrected Akaike information criterion.
 \end{flushleft}
\end{table*}

To start with, I fit the \textit{Suzaku} and \textit{Swift}/BAT
spectra of 3C~345 with a simple broken power law model, which is represented as
\textbf{zphabs*zbknpower}\footnote{A local model and the redshift variant of \textbf{bknpower}.}
in XSPEC terminology and whose photon spectrum $F \left( E \right)$
is expressed as
\begin{equation}
 F \left( E \right) =
 \begin{cases}
  A E^{-\Gamma_{\textrm{soft}}} & E \le E_{\textrm{break}} \\
  A E^{\Gamma_{\textrm{hard}} - \Gamma_{\textrm{soft}}}_{\textrm{break}} \left( \dfrac{E}{1 \ \textrm{keV}} \right)^{-\Gamma_{\textrm{hard}}} & E > E_{\textrm{break}},
 \end{cases}
\end{equation}
where $E$ is the photon energy in the rest frame, $E_{\textrm{break}}$ is
the break energy of the power law component in the rest frame,
$\Gamma_{\textrm{soft}}$ and $\Gamma_{\textrm{hard}}$ are the photon indexes
for the soft and hard components, respectively,
based on \citet[][hereafter G03]{Gambill2003} and \citet[][hereafter B06]{Belsole2006};
the authors analysed the spectra of the core component of 3C~345 obtained
with \textit{Chandra} with this model.
For the fitting algorithm, the standard Levenberg-Marquardt method
(\texttt{leven} in XSPEC) is applied through this section.

Table~\ref{table-model-a} shows the best-fitting parameters obtained with Model A.
B06, for example, reported flattening of the 3C~345 spectrum
(the photon index of $\Gamma = 1.3$) above 1.7 keV (in the observer frame)
and a necessity of two power law components.
However, the simultaneous fit of the XISs, HXD/PIN, \textit{Swift}/BAT
spectra yields a quite different break energy
($E_{\textrm{break}} = 8 \ \textrm{keV}$) from those of G03 and B06.

For further investigation, the spectra are divided at 6 keV, which corresponds
to 9.6 keV ($\sim E_{\textrm{break}}$) in the source redshift, and fitted with
Model A separately.
The spectral fitting in the 0.5--6.0 keV band agrees with the results of G03
and B06 within the errors except for the photon index for the harder spectrum
($\Gamma_{\textrm{hard}}$);
in the 6.0--100 keV spectra, $\Gamma_{\textrm{hard}}$ and $E_{\textrm{break}}$
agree with the literature within the errors, while the photon index for
the softer spectrum is fixed at $\Gamma_{\textrm{soft}} = 1.7$ due to its very
weak constraint.
Note that the choice of $\Gamma_{\textrm{soft}}$ does not affect the other
best-fitting parameters.
Interestingly, the spectra in the 6.0--100 keV band suggest that this source can
be a CT-AGN.
In such case, the transmitted component of the intrinsic power law through
the absorber is strongly suppressed due to the strong photoelectric absorption,
and the ``scattered'' (unabsorbed) and Compton reflection ones are only detectable
below 10 keV;
the transmitted component becomes comparable to the scattered one around this
energy, and the spectrum seems to be flatter there.
Previous studies below 10 keV could observe only the unabsorbed scattered
component.

\section{Torus Absorption Models} \label{section-torus-absorption}

\subsection{Analytic Torus Model}

Since the best-fitting parameters with Model A suggest that 3C~345
can be a CT-AGN,
I fit the spectra with an absorbed power law model
with an exponential cutoff, an absorbed Compton reflection and its reprocessed lines
from an infinitely thick reflector (Model B), which is often applied to
obscured AGNs \citep[e.g.,][]{Eguchi2009}.
Model B is written as \textbf{zphabs*zhighect*zpowerlw + const*zhighect*zpowerlw + zphabs*pexmon}
in XSPEC terminology, and the photon spectrum is represented as
\begin{eqnarray}
 F \left( E \right) = & \exp \left\{ - N_{\textrm{H}} \sigma \left( E \right) \right\} I \left( E \right) \notag \\
                      & + \exp \left\{ - N_{\textrm{H}}^{\textrm{refl}} \sigma \left( E \right) \right\} C \left( E \right) \notag \\
                      & + f I \left( E \right), \label{eq-model-b}
\end{eqnarray}
where $I \left( E \right) \equiv A E^{- \Gamma} \exp \left( - E / E_{\textrm{cut}} \right)$
is the intrinsic cutoff power law component, $\Gamma$ is the photon index,
$E_{\textrm{cut}}$ is the cutoff energy of the power law component,
$N_{\textrm{H}}$ is the line-of-sight hydrogen column density of absorbed component,
$\sigma \left( E \right)$ is the cross-section of photoelectric absorption,
$C \left( E \right)$ is the Compton reflection component \citep{Magdziarz1995}
and its reprocessed lines \citep{Nandra2007},
$N_{\textrm{H}}^{\textrm{refl}}$ is the line-of-sight hydrogen column density
of the reflection component, and $f$, which corresponds to the scattered fraction
in type 2 AGNs, is the relative strength of unabsorbed component with respect
to the intrinsic cutoff power law component.
A restriction of $N_{\textrm{H}} , N_{\textrm{H}}^{\textrm{refl}} \le 5 \times 10^{24} \ \textrm{cm}^{-2}$
is put on.
The inclination angle of the reflector viewed from the nucleus is
fixed at $60\degr$.
I let $E_{\textrm{cut}}$ free to vary within $0 \le E_{\textrm{cut}} \le 10^{4} \ \textrm{keV}$. 
The photon index, cutoff energy, and normalization of the unabsorbed and reflection components
are tied to those of the absorbed component.
The standard Levenberg-Marquardt method is applied here for the fitting algorithm.

\begin{table*}
 \centering
 \caption{
  Best-fitting parameters with Model B, E$_{1}$ and E$_{2}$
 }
 \label{table-model-be}
 \begin{tabular}{ccccc}
  \hline
   & Parameter & Model B & Model E$_{1}$ & Model E$_{2}$ \\
  \hline
  (1) & $N_{\textrm{H}}$ ($10^{24} \ \textrm{cm}^{-2}$) & 3.1 ($> 1.2$) & 3.3 ($> 1.8$) & $0^{a}$ \\
  (2) & $\Gamma$ & $1.68^{+0.06}_{-0.08}$ & $1.69^{+0.07}_{-0.09}$ & $1.69 \pm 0.09$ \\
  (3) & $E_{\textrm{cut}}$ (keV) & $> 129$ & $> 122$ & $> 267$ \\
  (4) & $N_{\textrm{H}}^{\textrm{refl}}$ ($10^{24} \ \textrm{cm}^{-2}$) & $= N_{\textrm{H}}$ & $0^{a}$ & $0^{a}$ \\
  (5) & $R$ & $8.6 \times 10^{-2}$ ($< 2.5$) & $1.9^{+11.3}_{-1.6} \times 10^{-2}$ & 0.33 ($< 0.81$) \\
  (6) & $f$ or $f_{c}$ (\%) & 37 ($> 24$) & $76^{+9}_{-11}$ & $100^{a}$ \\
  (7) & $A_{\textrm{BAT}}$ & $0.5^{+0.3}_{-0.1}$ & $0.5^{+0.4}_{-0.2}$ & $1.3^{+0.7}_{-0.4}$ \\
  (8) & $F_{\textrm{2-10}}$ ($10^{-12} \ \textrm{erg} \ \textrm{cm}^{-2} \ \textrm{s}^{-1}$) & 2.9 & 2.9 & 2.7 \\
  (9) & $F_{\textrm{10-50}}$ ($10^{-11} \ \textrm{erg} \ \textrm{cm}^{-2} \ \textrm{s}^{-1}$) & 1.5 & 1.5 & 0.59 \\
  (10) & $L_{\textrm{2-10}}$ ($10^{46} \ \textrm{erg} \ \textrm{s}^{-1}$) & 0.85 & 1.3 & 0.30 \\
  (11) & $\textrm{AIC}_{c}$ & $-85.19$ & $-85.33$ & $-76.50$ \\
   & $\chi^{2} / \textrm{d.o.f.}$ & $105.69 / 176$ & $105.61 / 176$ & $113.45 / 178$ \\
  \hline
 \end{tabular}
 \begin{flushleft}
  $^{a}$Fixed.
  \\
  Note.
  (1) The line-of-sight hydrogen column density of the absorbed power law component.
  (2) The power law photon index.
  (3) The cutoff energy of the intrinsic power law component.
  (4) The line-of-sight hydrogen column density of the reflection component.
  (5) The relative strength of the reflection component with respect to
      the intrinsic power law component, defined as $R \equiv \Omega / 2 \pi$,
      where $\Omega$ is the solud angle of the reflector viewed from the nucleus.
  (6) The relative strength of the unabsorbed (scattered) component
      with respect to the intrinsic power law component ($f$),
      or the covering fraction of the absorber ($f_{c}$).
  (7) The relative normalization of the BAT with respect to the FI-XISs.
  (8) The observed flux in the 2--10 keV band.
  (9) The observed flux in the 10--50 keV band.
  (10) The 2--10 keV intrinsic luminosity corrected for the absorption.
  (11) The corrected Akaike information criterion.
 \end{flushleft}
\end{table*}

Table~\ref{table-model-be} represents the best-fitting parameters obtained
with Model B.
The hydrogen column density of the reflection component $N_{\textrm{H}}^{\textrm{refl}}$
is linked to that of the absorbed power law one $N_{\textrm{H}}$
since $N_{\textrm{H}}^{\textrm{refl}}$ exceeds $N_{\textrm{H}}$
if there is no constraint.
The best-fitting value $N_{\textrm{H}} = 10^{24.5} \ \textrm{cm}^{-2}$
suggests that this source is a CT-AGN.
Note that only photoelectric absorption is considered, and that
Compton scattering, which cannot be negligible for a CT-AGN,
is not taken into account here;
fully consistent treatment is given in Section \ref{subsection-torus-model}.
The relative strength of the Compton reflection component with respect
to the intrinsic power law one $R$ is defined as $R \equiv \Omega / 2 \pi$,
where $\Omega$ is the solid angle of the reflector, and its best-fitting
value is remarkably small ($R = 8.6 \times 10^{-2}$) while its 90\%
upper limit also permits the reflection dominant case ($R < 2.5$).
On the other hand, the ``scattered'' fraction $f$ is significantly
larger than those of optical selected Seyfert 2 galaxies:
3--10\% \citep{Guainazzi2005}.
This can reflect a contamination by the jet components in 3C~345
since this source is a radio-loud AGN.

\subsection{Comparison of Model A and B}

The best-fitting $\chi^{2}$ value of Model B is less than that of Model A,
$\Delta \chi^{2} = - 5.88$.
To quantify whether this improvement is statistically significant,
I introduce Akaike information criterion \citep[AIC,][]{Akaike1974}.
The AIC is defined as
\begin{equation}
 \textrm{AIC} \equiv - 2 \ln \mathcal{L}_{\textrm{max}} + 2 k,
 \label{eq-aic-def}
\end{equation}
where $\mathcal{L}_{\textrm{max}}$ is the maximum likelihood
achievable by the model, and $k$ is the number of parameters
of the model.
The best model minimizes the AIC.
For $\chi^{2}$ minimization regime, Equation (\ref{eq-aic-def}) is
written as
\begin{equation}
 \textrm{AIC} = N \ln \dfrac{\chi^{2}_{\textrm{min}}}{N} + 2k,
\end{equation}
where $N$ is the number of data points \citep{Burnham2004}.
Since the AIC supposes that $N$ is infinite, a correction term is required
for a small sample size:
\begin{equation}
 \textrm{AIC}_{c} = \textrm{AIC} + \dfrac{2 k \left( k + 1 \right)}{N - k - 1}
\end{equation}
\citep{Sugiura1978}.
I denote the $\textrm{AIC}_{c}$ of the $i$-th model as
$\textrm{AIC}_{c, i}$, and define $\Delta_{i}$ as
\begin{equation}
 \Delta_{i} \equiv \textrm{AIC}_{c, i} - \textrm{AIC}_{c, \textrm{min}},
\end{equation}
where the subscript ``$\textrm{min}$'' represents the model
whose $\textrm{AIC}_{c}$ is smallest of the models.
An Akaike weight is defined as
\begin{equation}
 w_{i} \equiv \dfrac{\exp \left( - \Delta_{i} / 2 \right)}{\displaystyle \sum_{r = 1}^{R} \exp \left( - \Delta_{r} / 2 \right)},
\end{equation}
where $R$ is the number of the models, and this can be interpreted
as a model likelihood \citep{Akaike1981,Burnham2004}

I compute the $\textrm{AIC}_{c}$s for Model A and B, which are given
in Table~\ref{table-model-a} and \ref{table-model-be}, respectively;
since $\Delta_{A} = 7.8$ is obtained, Model B is not ``decisively''
but ``strongly'' preferred according to \citet{Liddle2007},
and their Akaike weights ($w_{B} / w_{A}$) suggest that the odds ratio is
approximately 50:1 against Model A.
Hence I conclude that Model B is better than Model A.

\subsection{Numerical Torus Model} \label{subsection-torus-model}

\begin{figure*}
 \centerline{
	\includegraphics[width=0.9\columnwidth]{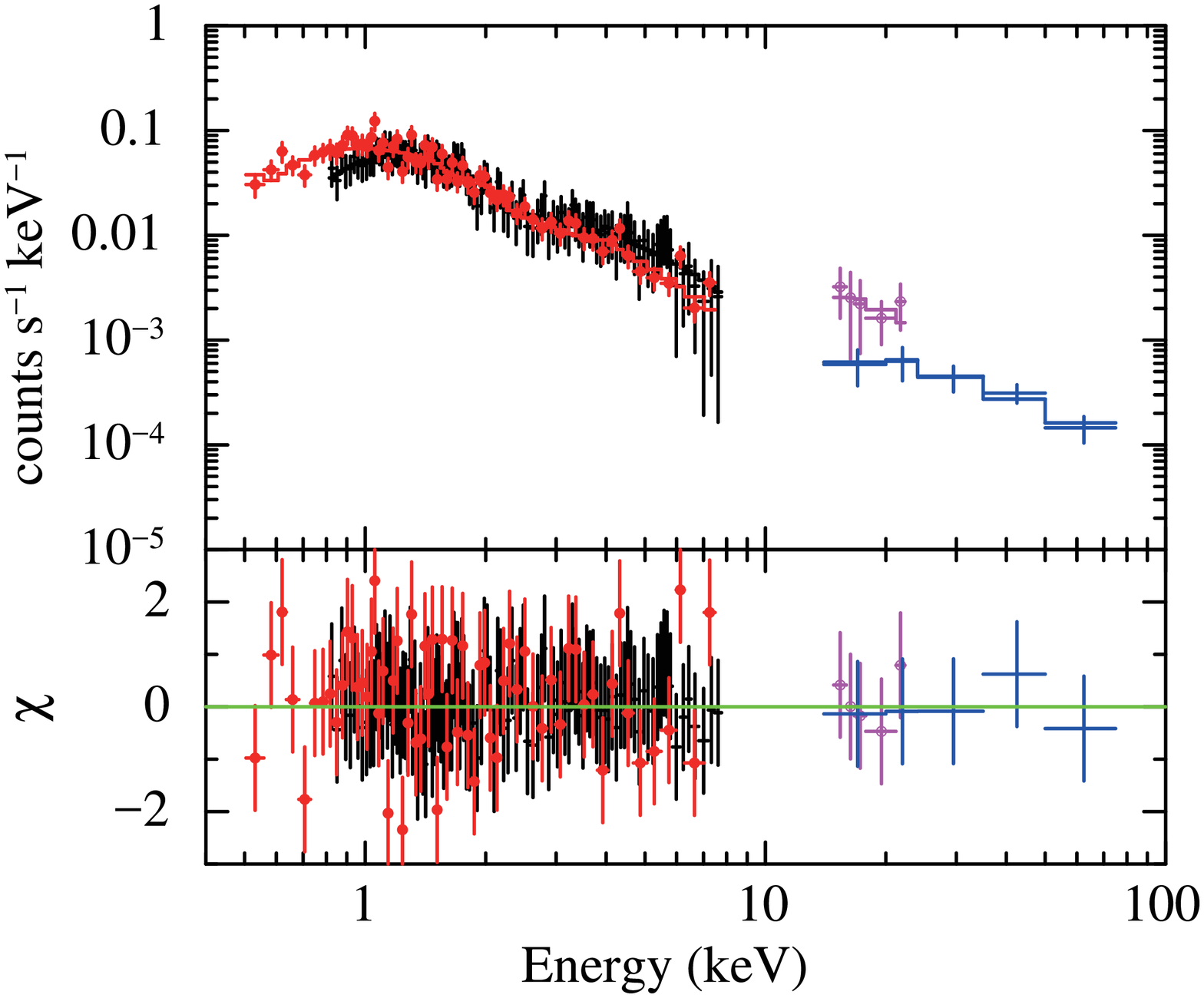}
	\hspace{0.1\columnwidth}
	\includegraphics[width=0.9\columnwidth]{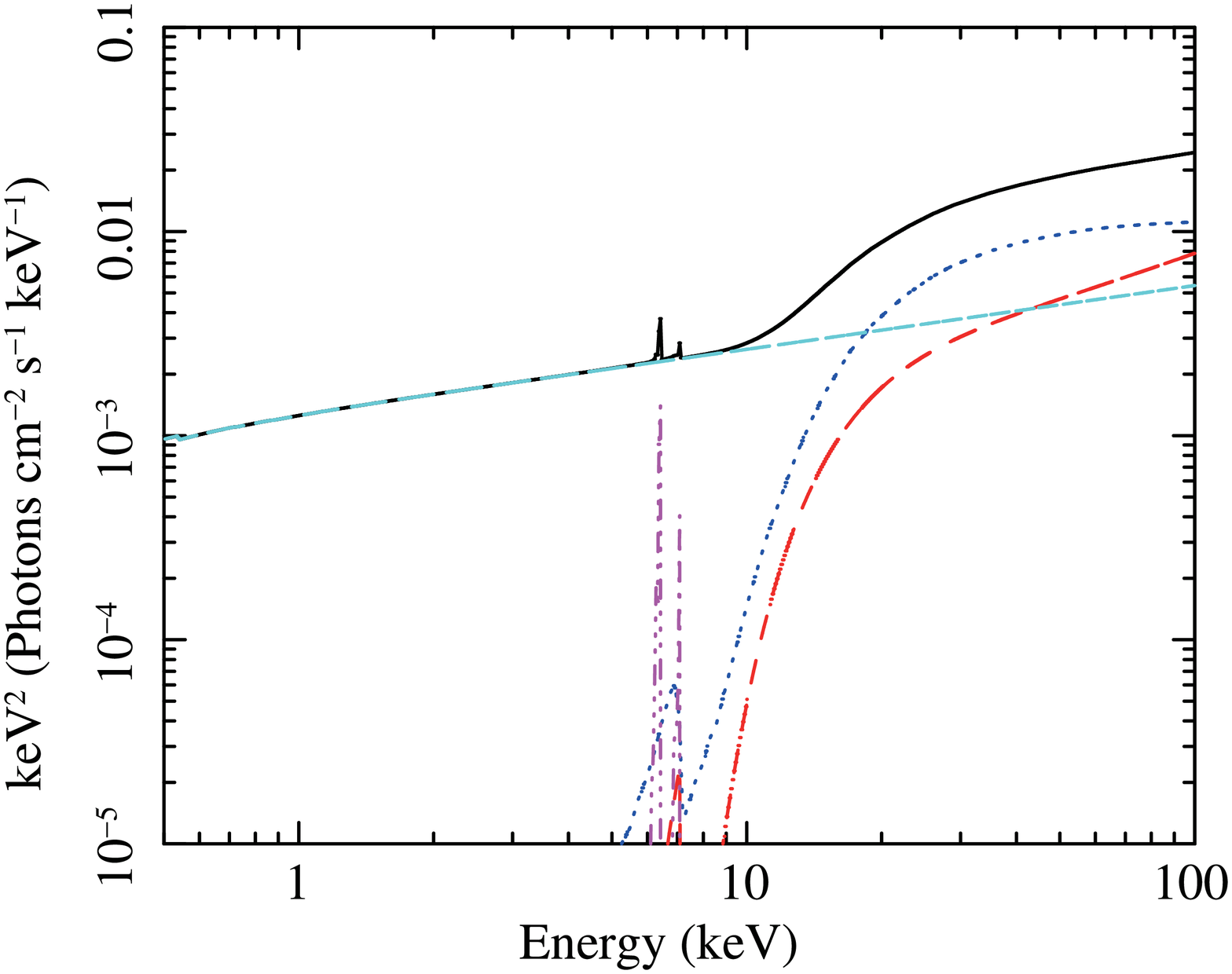}
 }
 \caption{
  Observed spectra (left) and the best-fitting spectral model (right) with
  Model C.
  Left: the black crosses, red filled circles, magenta open circles,
  and blue crosses correspond to the data of the FI-XISs,
  BI-XIS, HXD/PIN, and BAT, respectively, with their $1\sigma$ error bars.
  The spectra of the XIS and PIN are folded with the detector response
  in units of $\textrm{counts} \ \textrm{s}^{-1} \ \textrm{keV}^{-1}$,
  while those of the BAT are corrected for the detector area and
  have units of $\textrm{photons} \ \textrm{cm}^{-2}
  \ \textrm{ks}^{-1} \ \textrm{keV}^{-1}$.
  The best-fitting results are plotted by solid lines, and the residuals
  in units of $\chi$ are shown in the lower panels.
  Right: the best-fitting spectral model in units of $E F_{E}$ (where $E$ is
  the photon energy in the rest frame and $F_{E}$ is the photon spectrum);
  the solid black, dashed red, dotted blue, dotted-dashed cyan,
  dotted-dotted-dashed magenta curves correspond to the total,
  absorbed transmitted component, absorbed reflected component,
  unabsorbed power law component, and emission lines, respectively.
 }
 \label{figure-spectrum-model-c}
\end{figure*}

\begin{table}
 \centering
 \caption{
  Best-fitting parameters with Model C
 }
 \label{table-model-c}
 \begin{tabular}{ccc}
  \hline
  & Parameter & Model C \\
  \hline
  (1) & $N_{\textrm{H}}^{\textrm{Eq}}$ ($10^{24} \ \textrm{cm}^{-2}$) & $3.2 \pm 0.9$ \\
  (2) & $\theta_{\textrm{inc}}$ (degree) & $88 \pm 2$ \\
  (3) & $\Gamma$ & $1.69 \pm 0.05$ \\
  (4) & $f$ (\%) & $10 \pm 1$ \\
  (5) & $A_{\textrm{BAT}}$ & $0.47 \pm 0.08$ \\
  (6) & $F_{\textrm{2-10}}$ ($10^{-12} \ \textrm{erg} \ \textrm{cm}^{-2} \ \textrm{s}^{-1}$) & 2.8 \\
  (7) & $F_{\textrm{10-50}}$ ($10^{-11} \ \textrm{erg} \ \textrm{cm}^{-2} \ \textrm{s}^{-1}$) & 1.5 \\
  (8) & $L_{\textrm{2-10}}$ ($10^{46} \ \textrm{erg} \ \textrm{s}^{-1}$) & 3.1 \\
  (9) & $\textrm{AIC}_{c}$ & $-86.70$ \\
  & $\chi^{2} / \textrm{d.o.f.}$ & $106.08 / 177$ \\
  \hline
 \end{tabular}
 \begin{flushleft}
  (1) The hydrogen column density of the torus viewed from the equatorial direction.
  (2) The inclination angle of the torus.
  (3) The power law photon index.
  (4) The fraction of unabsorbed component relative to absorbed one.
  (5) The relative normalization of the BAT with respect to the FI-XISs.
  (6) The observed flux in the 2--10 keV band.
  (7) The observed flux in the 10--50 keV band.
  (8) The 2--10 keV intrinsic luminosity corrected for the absorption.
  (9) The corrected Akaike information criterion.
 \end{flushleft}
\end{table}

MY09 performed Monte Carlo simulations to obtain X-ray spectra from
a toroidal torus with the half-opening angle of the torus of $60\degr$.
The results are distributed as a set of FITS tables\footnote{\url{http://www.mytorus.com/}},
and can be imported via the \textbf{atable} and \textbf{etable}
models in XSPEC.
I refer to the results as MYTorus model.

In MYTorus model, Thomson and relativistic Compton scattering processes
are taken into account in addition to photoelectric absorption.
Thus MYTorus model is much more reliable than Model B even for CT cases.
MYTorus model originally consists of three components: MYTZ, MYTS, and MYTL.
MYTZ is the zeroth-oder continuum, that is, absorbed transmitted power law
component through the torus, MYTS is a sum of the absorbed and unabsorbed
reflected continua by the torus, and MYTL represents the reprocessed emission
lines by the torus.
MYTS and MYTL have three parameters: the photon index $\Gamma$,
the hydrogen column density of the torus viewed from the equatorial
direction $N_{\textrm{H}}^{\textrm{Eq}}$, and the inclination angle
of the torus $\theta_{\textrm{inc}}$ while MYTZ has two parameters:
$N_{\textrm{H}}^{\textrm{Eq}}$ and $\theta_{\textrm{inc}}$.

In this subsection, I apply MYTorus model to the \textit{Suzaku}
and \textit{Swift}/BAT spectra of 3C~345 to draw the physical
property of its absorber, which is temporarily assumed to be a dust
torus here.
I define Model C as a sum of these three torus components together
with an ``scattered'' unabsorbed power law component;
the photon spectrum $F \left( E \right)$ of Model C can be written as
\begin{eqnarray}
 F \left( E \right) = & \exp \left\{ - \textrm{MYTZ} \left( N_{\textrm{H}}^{\textrm{Eq}}, \theta_{\textrm{inc}}, E \right) \right\} I^{\prime} \left( E \right) \notag \\
                      & + \textrm{MYTS} \left( N_{\textrm{H}}^{\textrm{Eq}}, \theta_{\textrm{inc}}, \Gamma, E \right) \notag \\
                      & + \textrm{MYTL} \left( N_{\textrm{H}}^{\textrm{Eq}}, \theta_{\textrm{inc}}, \Gamma, E \right) \notag \\
                      & + f I^{\prime} \left( E \right),
\end{eqnarray}
where $I^{\prime} \left( E \right) \equiv A E^{- \Gamma}$ is the intrinsic
power law component, $E$ is the photon energy in the rest frame,
and $f$ is the relative strength of the unabsorbed power law component
with respect to the intrinsic power law one.
Model C is also written as 
\textbf{etable\{mytorus\_Ezero\_v00.fits\}*zpowerlw + atable\{mytorus\_scatteredH500\_v00.fits\} + atable\{mytl\_V000010nEp000H500\_v00.fits\} + const*zpowerlw}
in XSPEC terminology.
For the fitting algorithm, the MINUIT MIGRAD method (\texttt{migrad}
in XSPEC) is applied through this subsection since the standard 
Levenberg-Marquardt method is sometimes inappropriate for table models.

The resultant best-fitting parameters of Model C are summarized in
Table~\ref{table-model-c}.
We find that we are seeing the absorber (torus) of this object from
a completely edge-on angle ($\theta_{\textrm{inc}} \simeq 90\degr$).
The hydrogen column density ($N_{\textrm{H}}^{\textrm{Eq}} = 10^{24.5} \ \textrm{cm}^{-2}$)
and the photon index $\Gamma$ are consistent with those obtained with
Model B, suggesting that this source is a CT-AGN.
On the other hand, the strength of the unabsorbed power law component
relative to the intrinsic one is about one third of Model B.

The observed spectra fitted with Model C and the model spectrum are
shown in Figure~\ref{figure-spectrum-model-c}.
CT-AGNs usually show a distinct Fe K$\alpha$ emission and deep Fe K
edge in their X-ray spectra, but we cannot see such features in our
spectra.
The simple explanation for this is that the spectra below 10 keV
are completely dominated by the unabsorbed power law component
as can be seen from the figures, and it completely hides
such spectral features.

\section{Partial Covering Absorber Models} \label{section-partial-covering}

Recent studies find that in some broad line Seyfert 1 galaxies their
nuclei are partially covered with CT absorbers \citep[e.g.,][]{Turner2009}.
Since the line-of-sight hydrogen column density of 3C~345 is extremely high,
I investigate whether partial covering absorber models can account for
the spectra of 3C~345 in this section.

\subsection{Cloud-Like Absorber Model}

\begin{table}
 \centering
 \caption{
  Best-fitting parameters with Model D
 }
 \label{table-model-d}
 \begin{tabular}{ccc}
  \hline
   & Parameter & Model D \\
  \hline
  (1) & $N_{\textrm{H}}$ ($10^{24} \ \textrm{cm}^{-2}$) & $2.4 \pm 0.6$ \\
  (2) & $\Gamma$ & $1.69 \pm 0.05$ \\
  (3) & $f_{c}$ (\%) & $94 \pm 2$ \\
  (4) & $A_{\textrm{BAT}}$ & $0.50 \pm 0.09$ \\
  (5) & $F_{\textrm{2-10}}$ ($10^{-12} \ \textrm{erg} \ \textrm{cm}^{-2} \ \textrm{s}^{-1}$) & 3.0 \\
  (6) & $F_{\textrm{10-50}}$ ($10^{-11} \ \textrm{erg} \ \textrm{cm}^{-2} \ \textrm{s}^{-1}$) & 1.4 \\
  (7) & $L_{\textrm{2-10}}$ ($10^{46} \ \textrm{erg} \ \textrm{s}^{-1}$) & 5.1 \\
  (8) & $\textrm{AIC}_{c}$ & $-88.67$ \\
   & $\chi^{2} / \textrm{d.o.f.}$ & $106.19 / 178$ \\
  \hline
 \end{tabular}
 \begin{flushleft}
  (1) The line-of-sight hydrogen column density of the absorber.
  (2) The power law photon index.
  (3) The covering fraction of the absorber.
  (4) The relative normalization of the BAT with respect to the FI-XISs.
  (5) The observed flux in the 2--10 keV band.
  (6) The observed flux in the 10--50 keV band.
  (7) The 2--10 keV intrinsic luminosity corrected for the absorption.
  (8) The corrected Akaike information criterion.
 \end{flushleft}
\end{table}

I consider the case that a CT cloud, not a torus,
is incidentally in our line of sight.
This can be approximated by the zeroth-oder continuum represented
by $\textrm{MYTZ}$ in MYTorus model \citep{Yaqoob2012}.
I define Model D as follows:
\begin{eqnarray}
 F \left( E \right) = & f_{c} \exp \left\{ - \textrm{MYTZ} \left( N_{\textrm{H}}, \theta_{\textrm{inc}} = 90\degr, E \right) \right\} I^{\prime} \left( E \right) \notag \\
                      & + \left( 1 - f_{c} \right) I^{\prime} \left( E \right),
\end{eqnarray}
where $f_{c}$ corresponds to the covering fraction of the cloud.
Model D can be expressed as
\textbf{const*etable\{mytorus\_Ezero\_v00.fits\}*zpowerlw + const*zpowerlw}
in XSPEC terminology.

The best-fitting parameters of Model D are summarized in Table~\ref{table-model-d}.
The line-of-sight hydrogen column density is
$N_{\textrm{H}} = 10^{24.4} \ \textrm{cm}^{-2}$, and the covering
fraction is $f_{c} \simeq 95\%$, meaning that a CT absorber covers
a large solid angle of the nucleus, and it is likely to be not
a cloud but a classical torus.

\subsection{Compton Reflection Model}

As is obvious from Figure~\ref{figure-spectrum-model-c},
the flux above 20 keV of 3C~345 is remarkably higher than
that below 20 keV.
Since such spectral feature can be explained by a strong Compton
reflection component, Model E$_{1}$ is defined as follows:
\begin{eqnarray}
F \left( E \right) = & f_{c} \exp \left\{ - N_{\textrm{H}} \sigma \left( E \right) \right\} I \left( E \right) \notag \\
                     & + \exp \left\{ - N_{\textrm{H}}^{\textrm{refl}} \sigma \left( E \right) \right\} C \left( E \right) \notag \\
                     & + \left( 1 - f_{c} \right) I \left( E \right). \label{eq-model-e}
\end{eqnarray}
This model is written as \textbf{zpcfabs*zhighect*zpowerlw + zphabs*pexmon}
in the XSPEC terminology.
Here $N_{\textrm{H}}^{\textrm{refl}}$ and the inclination angle
of the reflector are fixed at 0 and $60\degr$, respectively.
As compared to Equation~(\ref{eq-model-b}) and (\ref{eq-model-e}),
Model E$_{1}$ is mathematically same as Model B except for the coefficient
of the unabsorbed power law component $I \left( E \right)$ ($f \rightarrow \left( 1 - f_{c} \right)$).
Note that the reflector is assumed to be the accretion disc or
clouds in the BLR in Model E$_{1}$ while it is the torus in Model B.

The best-fitting parameters are summarized in Table~\ref{table-model-be}.
We find the relative strength of the Compton reflection component is 
very weak ($R < 0.13$).
This means that the spectral curvature around 20 keV is explained
not by the Compton reflection but by heavy absorption.
However, such spectral feature should be accounted for by
the reflection generally.
For further investigation, the hydrogen column density of the power law
component and the covering fraction of the absorber in Model E$_{1}$
are fixed at $N_{\textrm{H}} = 0$ and $f_{c} = 100\%$, respectively,
as an extreme case (Model E$_{2}$).
The best-fitting parameters are also summarized in Table~\ref{table-model-be};
while a relatively strong reflection component is permitted ($R < 0.81$),
the resultant $\chi^{2}$ and $\textrm{AIC}_{c}$ indicate
that Model E$_{2}$ is inferior to Model E$_{1}$.
Hence the reflection is unlikely to be essential to account for
the spectral shape of this source.

\subsection{Ionized Absorber Models}

Observations with hard X-ray detectors recently revealed that
there is a category called ``hard excess'' AGNs, where the X-ray flux
above 10 keV is unexpectedly stronger than that predicted from
the $< 10$-keV spectrum, in type 1 AGNs \citep[e.g.,][]{Walton2010, Tatum2013}.
Though the physics of hard excess AGNs is not well understood
at this time, their hard X-ray spectra cannot be explained by
Compton reflection models like \textbf{pexrav} but accounted for
(multi-zone) ionized absorbers possibly in the BLR.
Such models are investigated in this subsection.

\subsubsection{Warm Absorber Model}

\begin{table*}
 \centering
 \caption{
  Best-fitting parameters with Model F$_{1}$ and F$_{2}$
 }
 \label{table-model-f}
 \begin{tabular}{cccc}
  \hline
   & Parameter & Model F$_{1}$ & Model F$_{2}$ \\
  \hline
  (1) & $\Gamma$ & $1.69^{+0.08}_{-0.07}$ & $1.8 \pm 0.1$ \\
  (2) & $N_{\textrm{H}}$ ($10^{24} \ \textrm{cm}^{-2}$) & 3.8 ($> 2.0$) & 4.4 ($> 2.3$) \\
  (3) & $f_{c}$ (\%) & $76^{+8}_{-19}$ & $78^{+8}_{-18}$ \\
  (4) & $\log \xi$ ($\textrm{erg} \ \textrm{cm}^{-2} \ \textrm{s}^{-1}$) & 1.4 ($< 2.0$) & 1.3 ($< 2.1$) \\
  (5) & $T$ (K) & $1.5 \times 10^{5 \, a}$ & $1.5 \times 10^{5 \, a}$ \\
  (6) & $N_{\textrm{H}, 2}$ ($10^{24} \ \textrm{cm}^{-2}$) & --- & $0.1^{b}$ \\
  (7) & $f_{c, 2}$ (\%) & --- & 19 ($< 41$) \\
  (8) & $\log \xi_{2}$ ($\textrm{erg} \ \textrm{cm}^{-2} \ \textrm{s}^{-1}$) & --- & $-1.1$ ($< 1.9$) \\
  (9) & $T_{2}$ (K) & --- & $3 \times 10^{4 \, a}$ \\
  (10) & $A_{\textrm{BAT}}$ & $0.5^{+0.4}_{-0.2}$ & $0.5^{+0.4}_{-0.2}$ \\
  (11) & $F_{\textrm{2-10}}$ ($10^{-12} \ \textrm{erg} \ \textrm{cm}^{-2} \ \textrm{s}^{-1}$) & 3.0 & 3.0 \\
  (12) & $F_{\textrm{10-50}}$ ($10^{-11} \ \textrm{erg} \ \textrm{cm}^{-2} \ \textrm{s}^{-1}$) & 1.5 & 1.5 \\
  (13) & $L_{\textrm{2-10}}$ ($10^{46} \ \textrm{erg} \ \textrm{s}^{-1}$) & 1.3 & 1.5 \\
  (14) & $\textrm{AIC}_{c}$ & $-88.64$ & $-84.22$ \\
   & $\chi^{2} / \textrm{d.o.f.}$ & $104.97 / 177$ & $103.71 / 174$ \\
  \hline
 \end{tabular}
 \begin{flushleft}
  $^{a}$Fixed.
  \\
  $^{b}$The error cannot be constrained.
  \\
  (1) The power law photon index.
  (2) The line-of-sight hydrogen column density of the (primary) absorber.
  (3) The covering fraction of the (primary) absorber.
  (4) The ionization parameter of the (primary) absorber.
  (5) The temperature of the (primary) absorber.
  (6) The line-of-sight hydrogen column density of the secondary absorber.
  (7) The covering fraction of the secondary absorber.
  (8) The ionization parameter of the secondary absorber.
  (9) The temperature of the secondary absorber.
  (10) The relative normalization of the BAT with respect to the FI-XISs.
  (11) The observed flux in the 2--10 keV band.
  (12) The observed flux in the 10--50 keV band.
  (13) The 2--10 keV intrinsic luminosity corrected for the absorption.
  (14) The corrected Akaike information criterion.
 \end{flushleft}
\end{table*}

I fit the spectra with the \textbf{absori} model in XSPEC
\citep{Done1992}, which represents the absorption by a spherical
warm ionized absorber.
There are 4 parameters in the \textbf{absori} model: the photon
index $\Gamma$, the line-of-sight hydrogen column density
$N_{\textrm{H}}$, the temperature of the absorber $T$, and
the ionization parameter defined by $\xi = L_{\textrm{ion}} / n r^{2}$,
where $L_{\textrm{ion}}$ is the isotropic luminosity of
the ionization source, and $n$ is the gas density of the absorber
at a distance of $r$ from the centre.
Note that the \textbf{abosri} model does not take Compton scattering
into account.
The standard Levenberg-Marquardt method is applied for the fitting
algorithm here.

Firstly, I consider a one-zone absorber model (Model F$_{1}$)
which is described as follows:
\begin{equation}
 F \left( E \right) = f_{c} K \left( \Gamma, N_{\textrm{H}}, \xi, T \right) I^{\prime} \left( E \right) + \left( 1 - f_{c} \right) I^{\prime} \left( E \right),
\end{equation}
where $f_{c}$ is the covering fraction of the absorber, and
$K \left( \Gamma, N_{\textrm{H}}, \xi, T \right)$ corresponds
to the \textbf{absori} model.
This model can be written as \textbf{const*absori*zpowerlw + const*zpowerlw}
in the XSPEC terminology.
The temperature is fixed at $T = 1.5 \times 10^{5} \ \textrm{K}$
due to its very weak constraint.
The best-fitting parameters are summarized in Table~\ref{table-model-f};
we find that the hydrogen column density is notably high,
and that the absorber is not so ionized.

Next, I assume that the absorber consists of two zones
and secondary shell surrounds the primary spherical layer
(Model F$_{2}$).
The temperatures of the primary and secondary layers are fixed
at $T = 1.5 \times 10^{5} \ \textrm{K}$ and
$T_{2} = 3 \times 10^{4} \ \textrm{K}$, respectively.
The photon spectrum is written as
\begin{eqnarray}
 F \left( E \right) = & \left\{ f_{c, 2} K \left( \Gamma, N_{\textrm{H}, 2}, \xi_{2}, T_{2} \right) + \left( 1 - f_{c, 2} \right) \right\} \notag \\
                      & \times \left\{ f_{c} K \left( \Gamma, N_{\textrm{H}}, \xi, T \right) + \left( 1 - f_{c} \right) \right\} I^{\prime} \left( E \right),
\end{eqnarray}
where $N_{\textrm{H}}$, $\xi$, and $f_{c}$ correspond to
the line-of-sight hydrogen column density, the ionization parameter,
and the covering fraction of the primary layer, respectively,
and $N_{\textrm{H}, 2}$, $\xi_{2}$, and $f_{c, 2}$ are
those of the secondary layer.

The best-fitting parameters are also summarized in Table~\ref{table-model-f}.
As is obvious from this table, the secondary layer is physically
meaningless since $N_{\textrm{H}, 2} \ll N_{\textrm{H}}$ and
the covering fraction of it is small.
This is also supported by the comparison of the $\textrm{AIC}_{c}$
values;
Model F$_{1}$ is 9 times as strong as Model F$_{2}$.
Hence the one-zone absorber model (Model F$_{1}$) is adequate
for 3C~345.

\subsubsection{Ionization and Thermal Equilibrium Model}

\begin{figure*}
 \centerline{
  \includegraphics[width=0.9\columnwidth]{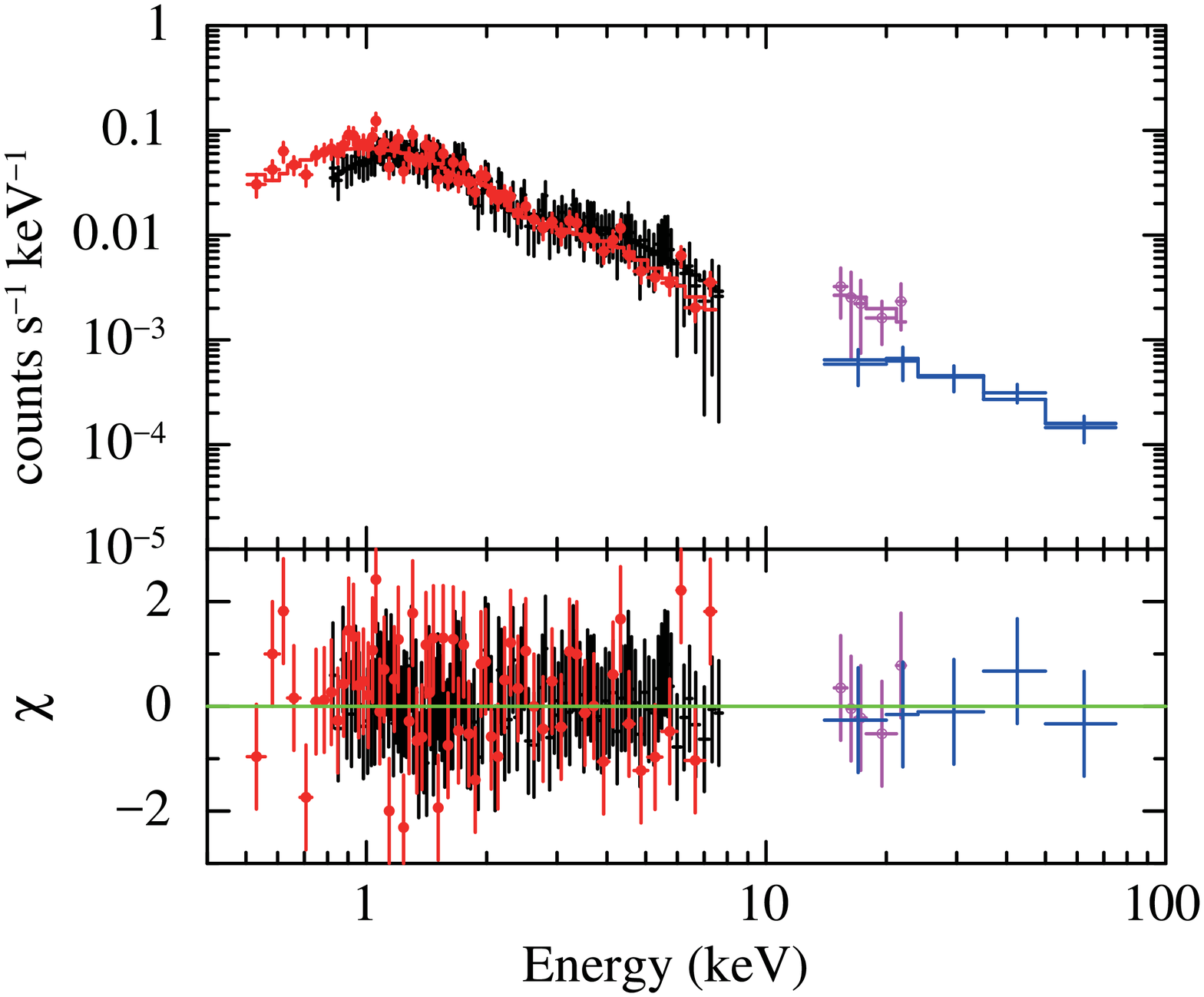}
  \hspace{0.1\columnwidth}
  \includegraphics[width=0.9\columnwidth]{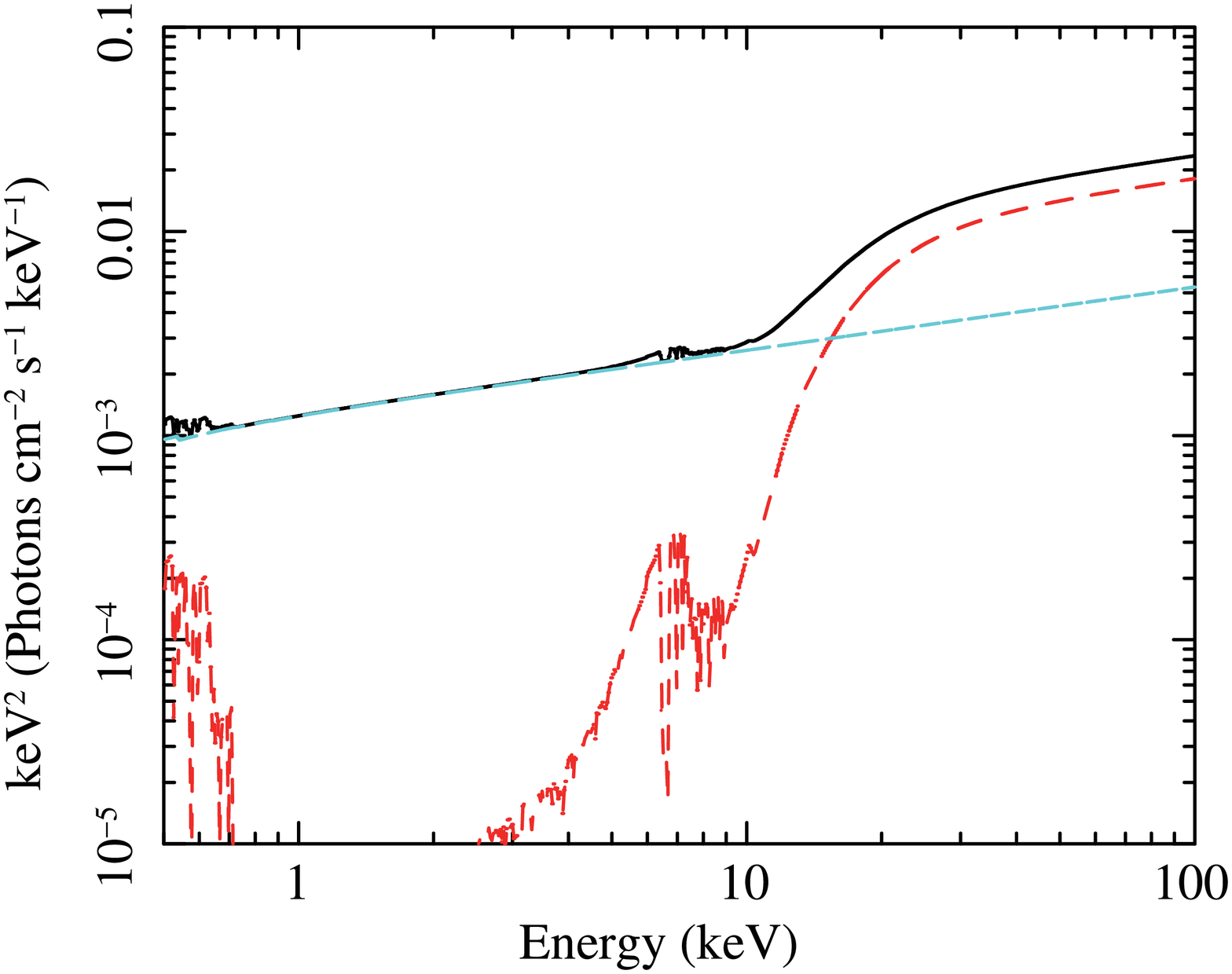}
 }
 \caption{
  Observed spectra (left) and the best-fitting spectral model (right) with
  Model G$_{10}$.
  Left: the black crosses, red filled circles, magenta open circles,
  and blue crosses correspond to the data of the FI-XISs,
  BI-XIS, HXD/PIN, and BAT, respectively, with their $1\sigma$ error bars.
  The spectra of the XIS and PIN are folded with the detector response
  in units of $\textrm{counts} \ \textrm{s}^{-1} \ \textrm{keV}^{-1}$,
  while those of the BAT are corrected for the detector area and
  have units of $\textrm{photons} \ \textrm{cm}^{-2}
  \ \textrm{ks}^{-1} \ \textrm{keV}^{-1}$.
  The best-fitting results are plotted by solid lines, and the residuals
  in units of $\chi$ are shown in the lower panels.
  Right: the best-fitting spectral model in units of $E F_{E}$ (where $E$ is
  the photon energy in the rest frame and $F_{E}$ is the photon spectrum);
  the solid black, dashed red, and dotted-dashed cyan curves correspond
  to the total, absorbed transmitted component, and unabsorbed power law
  component, respectively.
 }
 \label{figure-spectrum-model-g}
\end{figure*}

\begin{table*}
 \centering
 \caption{
  Best-fitting parameters with Model G$_{10}$, G$_{11}$, and G$_{12}$
 }
 \label{table-model-g}
 \begin{tabular}{ccccc}
  \hline
   & Parameter & Model G$_{10}$ & Model G$_{11}$ & Model G$_{12}$ \\
  \hline
  (1) & $n$ ($\textrm{cm}^{-3}$) & $10^{10 \, a}$ & $10^{11 \, a}$ & $10^{12 \, a}$ \\
  (2) & $\Gamma$ & $1.69^{a}$ & $1.69^{a}$ & $1.69^{a}$ \\
  (3) & $N_{\textrm{H}}$ ($10^{24} \ \textrm{cm}^{-2}$) & 9.8 ($> 9.2$) & 10 ($> 7.7$) & 9.8 ($> 8.5$) \\
  (4) & $f_{c}$ (\%) & $77.4 \pm 0.7$ & $79^{+3}_{-5}$ & $78.7 \pm 0.7$ \\
  (5) & $\log \xi$ ($\textrm{erg} \ \textrm{cm}^{-2} \ \textrm{s}^{-1}$) & $4 \pm 1$ & $1.95^{+0.14}_{-0.07}$ & $2.3 \pm 0.4$ \\
  (6) & $A_{\textrm{BAT}}$ & $0.48 \pm 0.08$ & $0.46^{+0.25}_{-0.09}$ & $0.46 \pm 0.08$ \\
  (7) & $F_{\textrm{2-10}}$ ($10^{-12} \ \textrm{erg} \ \textrm{cm}^{-2} \ \textrm{s}^{-1}$) & 2.9 & 2.8 & 2.8 \\
  (8) & $F_{\textrm{10-50}}$ ($10^{-11} \ \textrm{erg} \ \textrm{cm}^{-2} \ \textrm{s}^{-1}$) & 1.5 & 1.6 & 1.6 \\
  (9) & $L_{\textrm{2-10}}$ ($10^{46} \ \textrm{erg} \ \textrm{s}^{-1}$) & 1.3 & 1.4 & 1.4 \\
  (10) & $\textrm{AIC}_{c}$ & $-88.30$ & $-88.09$ & $-88.02$ \\
   & $\chi^{2} / \textrm{d.o.f.}$ & $105.16 / 177$ & $105.28 / 177$ & $105.32 / 177$ \\
  \hline
 \end{tabular}
 \begin{flushleft}
  $^{a}$Fixed.
  \\
  (1) The gas density of the absorber.
  (2) The power law photon index.
  (3) The line-of-sight hydrogen column density of the absorber.
  (4) The covering fraction of the absorber.
  (5) The ionization parameter of the absorber.
  (6) The relative normalization of the BAT with respect to the FI-XISs.
  (7) The observed flux in the 2--10 keV band.
  (8) The observed flux in the 10--50 keV band.
  (9) The 2--10 keV intrinsic luminosity corrected for the absorption.
  (10) The corrected Akaile information criterion.
 \end{flushleft}
\end{table*}

The limitations of the \textbf{absori} model are that
the cross-sections above 5 keV are simply approximated by $\propto E^{-3}$,
and that the absorber is in ionization equilibrium but not
in thermal equilibrium.
Though the spectra of 3C~345 do not show any clear line features,
I solve radiative transfer by utilizing the XSTAR code
to draw the more realistic nature of the absorber.
The version number of the code used here is 2.39, which
comes with the HEAsoft version 6.20 not with version 6.18,
since some fatal bugs were fixed in the HEAsoft version 6.19
and 6.20 (see their release notes for detail).
The MPI\_XSTAR program\footnote{\url{http://hea-www.cfa.harvard.edu/~adanehka/mpi_xstar/}}
is also employed for efficient usage of multi-core CPUs.

A spherical gas of uniform density is considered.
The gas is assumed to be ionized by the nucleus at the centre
with a single power law spectrum of a photon index of $\Gamma = 1.69$.
The covering fraction is fixed at 100\%, and let it free to vary
in XSPEC later by adding a normalization parameter $f_{c}$.
The turbulence velocity of the gas is set to be $300 \ \textrm{km} \ \textrm{s}^{-1}$.
Iterative calculations are performed until the gas is in
thermal equilibrium.
The results are compiled into three table models which XSPEC
can read by the \texttt{xstar2table} program;
only a table named \texttt{xout\_mtable.fits}, which contains
the absorption spectrum in the transmitted direction, is used
in this sub-subsection.
When this component is expressed as
$k_{n} \left( \Gamma, N_{\textrm{H}}, \xi \right)$
(the subscript ``$n$'' represents the density of the gas),
the photon spectrum of the model thought here is written as
\begin{equation}
 F \left( E \right) = f_{c} k_{n} \left( \Gamma, N_{\textrm{H}}, \xi \right) I^{\prime} \left( E \right) + \left( 1 - f_{c} \right) I^{\prime} \left( E \right),
\end{equation}
and \textbf{const*mtable\{xout\_mtable.fits\}*zpowerlw + const*zpowerlw}
in the XSPEC terminology.
Again, $\Gamma$ is fixed at 1.69.
Note that Compton scattering is not taken into account
in the XSTAR code.

The MINUIT MIGRAD method is applied here for the fitting algorithm.
Three cases of gas density are computed: $n = 10^{10}, 10^{11}, 10^{12} \ \textrm{cm}^{-3}$.
These are referred to as Model G$_{10}$, G$_{11}$, and G$_{12}$,
respectively.
The best-fitting parameters are summarized in Table~\ref{table-model-g}.
Figure~\ref{figure-spectrum-model-g} shows the model spectrum of
Model G$_{10}$.
Overall, the ionization parameters and the line-of-sight column
density of Model G$_{n}$ ($n = 10, 11, 12$) are larger than those of Model F$_{1}$.
The higher a gas is ionized, the more it becomes transparent since
electrons in atoms are more weakly bound and can move about freely.
Thus this tendency (a higher ionization parameter leads to
a higher column density) qualitatively seems correct.
On the other hand, the column densities are so high that
these models are not strictly accurate.

\section{Discussion} \label{section-discussion}

\subsection{Best-Fitting Models}

I fitted the observed spectra of 3C~345 with 7 different
models in this paper.
One of them (Model A) is a simple broken power model,
two of them (Model B and C) are the torus absorption models,
and four of them (Model D, E$_{x}$, F$_{x}$, G$_{n}$) are
the partial covering absorber models.
Since Model A has no advantage over the other models
due to its relatively large $\textrm{AIC}_{c}$ value,
it is not discussed below.

When Model B and C are compared, while the $\textrm{AIC}_{c}$
values are comparable, Model C is physically valid even
for CT cases.
Thus I choose Model C as a best-fitting model.
While Model D gives us the smallest $\textrm{AIC}_{c}$ of
the partial covering absorber models, the result that
a nearly full-coverage ($f_{c} \simeq 95\%$) absorber
is in our line of sight conflicts with the assumption
that the absorber is not a torus but a cloud.
In Model E$_{1}$, we expect that the spectral shape can be
accounted for by a strong Compton reflection component of
the accretion disc or clouds in the BLR, and that
the line-of-sight hydrogen column density should be rather
small.
However, the results lead to a very large column density
of $N_{\textrm{H}} = 10^{24.5} \ \textrm{cm}^{-2}$
and a very weak Compton reflection component of $R < 0.13$.
That is, the results conflict with the assumptions.
When $N_{\textrm{H}} = 0$ and $f_{c} = 100\%$ are
assumed (Model E$_{2}$), a rather strong reflection
component is allowed ($R < 0.81$).
However, this model is not favoured due to the large
$\textrm{AIC}_{c}$ value.
Though Model F$_{1}$ reproduces the shape of the observed
spectra well, there are some limitations due to
the \textbf{absori} model.
In that sense, Model G$_{n}$ are more appropriate than
Model F$_{1}$ since it takes radiative transfer into account
and the absorber is in ionization and thermal equilibrium
though Compton scattering process is not considered.
Thus I also choose Model G$_{n}$ as another best-fitting model.

\subsection{Impact of the Uncertainty of the HXD/PIN NXB Model}

As mentioned in Section \ref{subsection-data-reduction},
the NXB model for the HXD/PIN has systematic errors of
$\simeq 1.4\%$ at a $1\sigma$ confidence level in the
15--40 keV band for a 10 ks exposure \citep{Fukazawa2009},
which is almost same as the \textit{Suzaku} observation
of 3C~345.
To evaluate the impact of this uncertainty on my spectral
fitting, I create a NXB model where count rate in each energy
bin is gained 3\%, conservatively, from the original ``tuned''
NXB model.
This new NXB model is added to the CXB model spectrum used
in Section \ref{subsection-data-reduction}, and applied to
the HXD/PIN spectrum, which is simultaneously fitted with
Model C together with the XIS and \textit{Swift}/BAT spectra.
The resultant best-fitting parameters except for the relative
normalization of the \textit{Swift}/BAT with respect to
the FI-XISs $A_{\textrm{BAT}}$ fall within their original
90\% statistical errors;
$A_{\textrm{BAT}}$ changes from 0.47 to 0.55.
Similarly, I also create a NXB model whose count rate is
reduced by 3\%, and then apply it to the observed HXD/PIN
spectrum, this yielding consistent best-fitting parameters
with their original values within 90\% confidence limits. 

Likewise, the uncertainty of the HXD/PIN NXB model is
also investigated for Model G$_{n}$.
The covering fraction $f_{c}$ absolutely changes by 5\%
($77\% \rightarrow 82\%$, for example),
and the normalization of \textit{Swift}/BAT relative to
the FI-XISs $A_{\textrm{BAT}}$ also changes by 0.1.
The other parameters fall within the 90\% confidence limits of
their original values.
Hence I conclude that the systematic uncertainty of
the HXD/PIN NXB model does not affect our arguments.

\subsection{Interpretation of the Strong Unabsorbed Component of Model C}

The results fitted with Model C suggest that 3C~345 is a CT-AGN
with a strong scattered (unabsorbed) component of
$f = 10\%$, which corresponds to the upper limit of
the typical values of Seyfert 2 galaxies \citep[3--10\%,][]{Guainazzi2005}.
As pointed out in \citet{Brightman2014}, spectral fitting of
bright unobscured sources are sometimes misidentified
as CT-AGNs with a strong scattered component and a weak
underlying torus component.
Hence the authors introduced an upper limit of 10\% of scattered
component into their CT sample selection.
The $f$ value of Model C falls just on this border
line, and it is not rejected at this point.
A possible explanation to account for the strong scattered
component of this source is that the torus is more gaseous
and less dusty than typical type 2 AGNs, and incident photons
from the central engine are scattered in our direction
by the gas by Thomson scattering.
However, we have to accept the strange ``facts'' that 3C~345
is type 1 in the optical band but is type 2 in the X-ray band,
and that the torus is viewed from a completely edge-on angle
in addition.

\subsection{Bolometric Luminosity and Eddington Ratio}

The relative normalization of the \textit{Swift}/BAT
spectrum with respect to the FI-XISs is $A_{\textrm{BAT}} \simeq 0.5$
for all the models except for Model E$_{2}$, which is found
to be inappropriate for 3C~345.
That is, there is time variability between the \textit{Swift}/BAT
and \textit{Suzaku} observations, and the flux observed with
\textit{Suzaku} is twice as high as that extrapolated from
the \textit{Swift}/BAT spectrum.
Since the optical data are compared to the X-ray ones,
this correction is always considered below.

GCJ01 derived the supermassive black hole mass of 3C~345
of $\log \left( M_{\textrm{BH}} / M_{\sun} \right) = 9.901$ based
on the H$\beta$ line width.
S11 also derived the mass of $\log \left( M_{\textrm{BH}} / M_{\sun} \right) = 9.27 \pm 0.09$
and the Eddington ratio $\lambda_{\textrm{Edd}, \textrm{Opt}} = 0.79$ by utilizing
the optical H$\beta$ line width and its luminosity in the spectrum
of Sloan Digital Sky Survey (SDSS).
Both authors used the same indicator but obtained
different masses.
This can be due to their samples.
While GCJ01 focused on radio-loud quasars, S11 handled all SDSS
quasars.

Firstly, I discuss the Eddington ratio $\lambda_{\textrm{Edd}, \textrm{X}}$
derived from the X-ray luminosity with Model C.
The absorption and time variability corrected 2--10 keV band
luminosity is $L_{2-10} = 1.4 \times 10^{46} \ \textrm{erg} \ \textrm{s}^{-1}$.
The bolometric correction for the X-ray luminosity by \citet{Marconi2004}
yields a bolometric luminosity of
$L_{\textrm{bol}, \textrm{X}} = 1.4 \times 10^{48} \ \textrm{erg} \ \textrm{s}^{-1}$.
Since the Eddington luminosity based on S11 is
$L_{\textrm{Edd}} = 2.3 \times 10^{47} \ \textrm{erg} \ \textrm{s}^{-1}$,
the Eddington ratio derived from the X-ray luminosity is estimated
to be $\lambda_{\textrm{Edd}, \textrm{X}} = 5.9$ (super-Eddington),
6 times higher than that derived from the optical spectrum.
When the black hole mass derived by GCJ01 is applied,
the Eddington luminosity is $L_{\textrm{Edd}} = 1.0 \times 10^{48} \ \textrm{erg} \ \textrm{s}^{-1}$,
and the Eddington ratio is $\lambda_{\textrm{Edd}, \textrm{X}} = 1.4$.
Thus Model C is unlikely to explain the optical observations.

Next, I investigate Model G$_{10}$ as an example of Model G$_{n}$ similarly.
The 2--10 keV band luminosity is $L_{2-10} = 6.4 \times 10^{45} \ \textrm{erg} \ \textrm{s}^{-1}$,
and the bolometric luminosity is $L_{\textrm{bol}, \textrm{X}} = 5.9 \times 10^{47} \ \textrm{erg} \ \textrm{s}^{-1}$.
The Eddington ratio based on the black hole mass by S11 is
$\lambda_{\textrm{Edd}, \textrm{X}} = 2.5$.
However, $\lambda_{\textrm{Edd}, \textrm{X}} = \lambda_{\textrm{Edd}, \textrm{Opt}}$
yields a black hole mass of $\log \left ( M_{\textrm{BH}} / M_{\sun} \right) = 9.8$,
which is smaller than that by GCJ01.
Thus Model G$_{n}$ are likely to explain the optical spectrum.
When the fact that there is no Fe K$\alpha$ line and K absorption
edge in the X-ray spectrum of 3C~345 is also considered,
the best-fitting model for this source is likely to be Model G$_{n}$.

\subsection{Binary AGN Scenario}

A calculation of the 2--10 keV band luminosity of the unabsorbed
component in Model C yields $L_{2-10} = 1.4 \times 10^{45} \ \textrm{erg} \ \textrm{s}^{-1}$.
The bolometric correction for this luminosity by \citet{Marconi2004}
gives us $L_{\textrm{bol}, \textrm{X}} = 1.1 \times 10^{47} \ \textrm{erg} \ \textrm{s}^{-1}$.
When $\lambda_{\textrm{Edd}, \textrm{X}} = \lambda_{\textrm{Edd}, \textrm{Opt}}$
is assumed, we obtain the black hole mass of
$\log \left( M_{\textrm{BH}} / M_{\sun} \right) = 9.0$ or
$M_{\textrm{BH}} = 1.1 \times 10^{9} M_{\sun}$.
Interestingly, this value is slightly smaller than the lower
limit of the black hole mass by S11.

\citet{Lobanov2005} suggested that there is a supermassive
black hole binary with an equal mass of $M_{\textrm{BH}} = 7.1 \times 10^{8} M_{\sun}$
and the separation of $\sim 0.33 \ \textrm{pc}$ in 3C~345
based on the time variability in the optical and radio
bands and the precession of the jet.
The black hole mass obtained above is 1.5 times as heavy
as that by \citet{Lobanov2005}, but their scenario seems
attractive for Model C.
Let us assume that both black holes have their own accretion
discs and dust tori, and that one of them is
a heavily absorbed CT-AGN with a type 2 nucleus with
a completely edge-on viewing angle of the torus and
lies behind the other one with a type 1 nucleus.
An example of such systems is CID-42 \citep{Civano2010}.
The X-ray spectrum of the type 2 nucleus in the 2--10 keV
band is dominated by that of the type 1 nucleus due to
the strong photoelectric absorption, thus the sign of
the type 2 nucleus is missed in $\lesssim 10$-keV
observations, but detected in $\gtrsim 10$-keV observations.
This could be the case for the \textit{Suzaku} and \textit{Swift}/BAT
spectra, and could explain the strange nature that this source
is type 1 in the optical band but type 2 in the X-ray band,
and the torus is viewed from a completely edge-on angle.
Furthermore, if the type 2 nucleus belongs to ``hidden''
population \citep{Ueda2007, Winter2009}, it is hard
to detect it in the optical band since the flux of
the [\ion{O}{III}] line is too weak.

I have no evidence to prove this scenario at this time.
Even if this is the case, the confirmation is very
challenging even for future missions and telescopes.
However, a search for an offset [\ion{O}{III}] line
with respect to the source redshift could be worth doing.

\section{Summary} \label{section-summary}

The archival data of 3C~345 obtained with \textit{Suzaku}
and \textit{Swift}/BAT are analysed.
In previous studies, the X-ray spectra below 10 keV of
this source were fitted with a simple broken power law model
without absorption, but I found that the spectrum above 10 keV
is unexpectedly stronger than that predicted by the one below 10 keV.
Since such spectral shape can be explained by the strong
photoelectric absorption and Compton scattering in a dense
material generally, models for Compton thick AGNs and partial
covering absorbers in Seyfert 1 galaxies were applied to
the \textit{Suzaku} and \textit{Swift}/BAT spectra.

The numerical torus model by MY09, which represents the absorbed
transmitted component and reflection by the torus, suggests
that this source can be a Compton thick AGN with the hydrogen
column density of the torus of $N_{\textrm{H}}^{\textrm{Eq}} = 10^{24.5} \ \textrm{cm}^{-2}$,
the inclination angle of $\theta_{\textrm{inc}} \simeq 90\degr$,
and a relatively strong scattered component for typical
type 2 AGNs of $f = 10\%$.
However, the comparison of the Eddington ratio derived
from the 2--10 keV band luminosity to that of the SDSS
spectrum indicates that this source is shining at
a super Eddington luminosity, and this model seems inappropriate
except for the possibility that 3C~345 is a binary system of
supermassive black holes suggested by \citet{Lobanov2005}.

The partial covering ionized absorber model proposes that
this source is a hard excess AGN with the very large absorbing
column density of $N_{\textrm{H}} \simeq 10^{25} \ \textrm{cm}^{-2}$,
the ionization parameter of $\log \xi \gtrsim 2 \ \textrm{erg} \ \textrm{cm}^{-2} \ \textrm{s}^{-1}$,
and the covering fraction of $75\% \lesssim f_{c} \lesssim 85\%$.
Though the 2--10 keV band luminosity requires a relatively
large black hole mass of $\log \left( M_{\textrm{BH}} / M_{\sun} \right) = 9.8$,
which is heavier than that estimated from the SDSS spectrum,
but it is consistent with another optical observation.
Thus this model is likely to the best-fitting model for 3C~345.

To my knowledge, 3C~345 is the most distant and most absorbed
hard excess AGN.
Further detailed observations of this source at multi-wavelengths
would give us a deeper understanding of hard excess AGNs and
the cosmic evolution of supermassive black holes.

\section*{Acknowledgements}

I greatly thanks Kohei Ichikawa and Tessei Yoshida for
productive discussions, and appreciate Mikio Morii's
input on AIC.
I also appreciate insightful comments from the anonymous
referee.







\bsp	
\label{lastpage}
\end{document}